\documentclass[aps,twocolumn,showpacs]{revtex4}
\usepackage{graphicx}
\usepackage{bm}
\usepackage{amsmath}

\begin{document}

\title{Gossamer Superconductivity near Antiferromagnetic Mott Insulator in Layered Organic
Conductors}
\author{J. Y. Gan$^1$, Yan Chen$^2$, Z. B. Su$^{1}$, and F. C. Zhang$^{2,3,4}$}
\affiliation{$^1$Institute for Theoretical Physics, Chinese Academy of Sciences, Beijing,
100001, China \\
$^2$Department of Physics, University of Hong Kong, Pokfulam Road, Hong
Kong, China \\
$^3$Department of Physics, University of Cincinnati, Cincinnati, Ohio 45221\\
$^4$Department of Physics, Zhejiang University, Hangzhou, China}
\date{\today}

\begin{abstract}
Layered organic superconductors are on the verge of the Mott
insulator. We use Gutzwiller variational method to study a Hubbard
model including a spin exchange coupling term. The ground state is
found to be a Gossamer superconductor at small on-site Coulomb
repulsion $U$ and an antiferromagnetic Mott insulator at large $U$,
separated by a first order phase transition. Our theory is
qualitatively consistent with major experiments reported in organic
superconductors.

\end{abstract}

 \maketitle

\narrowtext

There has been much interest recently on the novel physics of
layered organic superconductor~\cite{organic,McKenzie97,
book1,john1,Lang03}. These compounds share most common physical
properties with the high-T$_{c}$ superconductor but
typically with much reduced temperature and energy scales.
$\kappa$-(BEDT-TTF)$_{2}$X
(X=anion) is a family of the best characterized organic
superconductors, where the quasi-2-dimensional (2D)
Fermi surface has been observed
and a direct first order transition between antiferromagnetic (AFM)
insulator and superconductor can be tuned by applied pressure or
magnetic fields~\cite{Miya95,Lef00,Miya02,Sasaki02,Muller02}. The
resemblance of its pressure-temperature phase diagram to that of the
carrier-density-temperature phase diagram in cuprates
and the fact of close proximity between the superconducting (SC) and
AFM insulating phases have been taken as evidences for similar
mechanisms governing high $T_c$ superconductors.
There have been strong evidences that the organic superconductors
are at the verge of the Mott insulator~\cite{Lef00,Miya95},
exhibiting the pseudogap phenomenon~\cite{Miya02}.
While an ongoing debate persists as to the precise symmetry of the singlet pairing,
more recent NMR~\cite{Maya95,Kanoda}, angular
dependent STM~\cite{arai} and thermal conductivity measurements in
the vortex state~\cite{izawa3} indicate a d$_{x^2-y^2}$ symmetry.

The low energy electronic structure of the organic superconductors
is well approximated by a 2D Hubbard model at the half filling.
Different from the cuprates, organic compounds can be SC at the half
filling, which makes $t$-$J$ model inappropriate to describe its SC
state. Most theoretical works so far have taken a weak-coupling
approach, in which a Hartree-Fock mean field ~\cite{Kino96},
fluctuation-exchange
approximation~\cite{Schmalian98,Kino98,Kondo99,Kino04} or random
phase approximation method ~\cite{Ogata04} are used. The weak
coupling theory gives a phase diagram of the AFM and SC states
qualitatively consistent with the experiments.  However, the weak
coupling theory has difficulties to address the Mott insulator or
the pseudogap phenomenon~\cite{Lef00,Miya95,Miya02}. The transition
between SC and AFM has also been investigated by using
renormalization group method~\cite{Nagaosa}.

Very recently, Laughlin has proposed a Gossamer Hamiltonian of which
a partially Gutzwiller projected BCS state is an exact ground state
with a tiny superfluid density at the half filling~\cite{laughlin}.
In that Hamiltonian, the SC state has an
instability toward the AFM ordering~\cite{laughlin2}. Some of the
present authors~\cite{zhang03,Gan03}
have examined the Gossamer superconductor, the
Mott insulator, and the resonating valence bond (RVB)
state~\cite{vannila,zhang88,randeria} in
strongly correlated electron systems with the
hope to unify the superconductivity in cuprates and in organic
compounds~\cite{coleman}. In our previous study, we focused on
the metallic/SC and insulating nature of the problem and neglected
the antiferromagnetism.
A related approach was recently
taken by Baskaran
~\cite{Baskaran}, who introduced a two-species $t$-$J$ model to describe
independent motions of empty sites and doubly occupied sites in an
otherwise spin-1/2 background, and discussed the relevance of the model
to  the organic superconductors.

In this Letter, we use Gutzwiller's variational method to study the
interplay between SC and AFM states in a modified Hubbard model in
2D given by Eqn. (1) below. By using a renormalized mean field
theory developed early for the $t$-$J$ model~\cite{zhang88}, we find
that at the half filling the ground state is an AFM Mott insulator
at large on-site repulsion $U$  and a Gossamer superconductor at
small $U$, followed by a normal metallic state at further smaller
$U$. The transition between the AFM and SC phases is first order,
and there is no co-existence of the two phases at the half filling.
The doping dependence of the model at large $U$ is similar to that
of the $t$-$J$ model~\cite{vannila}. Our results are qualitatively
consistent with major experiments in organic superconductors.

We consider a modified Hubbard model on a square lattice,
\begin{eqnarray}
H &=& U\sum_i n_{i\uparrow}n_{i\downarrow}- \sum\limits_{\langle ij \rangle\sigma}t_{ij}(c_{i\sigma}^\dagger c_{j\sigma}
+ h.c.)  \nonumber\\
   &&    + J \sum\limits_{(ij)} \vec{S}_i\cdot\vec{S}_j - \mu\sum_{i\sigma} n_{i\sigma}
\end{eqnarray}
In the above Hamiltonian, $c_{i\sigma}$ is an annihilation operator
of an electron at site $i$ with spin $\sigma$, $n_{i\sigma} =
c^{\dag}_{i\sigma}c_{\i\sigma}$, and $U >0$ is the on-site Coulomb
repulsion. The non-zero hopping integrals are $t_{ij} = t$ for the
nearest neighbor (n.n.)  pairs and $t_{ij} =t'$ for the next n.n.
pairs along [1,1] direction. $\vec S_i$ is a spin-1/2 operator, and
the summation in the spin exchange term is over all the n.n. pairs.
We consider $1 > t^\prime/t >0$, suitable for the organic compounds.
This Hamiltonian contains an additional spin exchange term to the
standard Hubbard model. In the limit $U \gg t$, the model is reduced
to the $t$-$t'$-$J$ model. At the half filling, the large $U$ limit
of the model is reduced to the AFM Heisenberg model with an AFM
ground state at small values of $t'/t$. At small $U$, we expect a
metallic or a SC ground state. We believe that the model combined
with the Gutzwiller trial wavefunction approach, Eqns. (2-3) below
is appropriate to
study the phase transitions in organic superconductors. Note that
the direct application of the Gutzwiller trial wavefunction
to the  Hubbard model is hardly to obtain the SC
pairing because of the non-explicit form of the spin-spin exchange
interaction in the Hamiltonian.

To study the phase transition between the AFM and SC states, we
consider a partially Gutzwiller projected spin density wave
(SDW)-BCS wavefunction~\cite{Lee,Gros},
\begin{eqnarray}
&&|\Psi_{GS} \rangle = \prod\limits_i (1 - \alpha n_{i\uparrow}n_{i\downarrow})|\Psi_{0}\rangle \\
&&|\Psi_{0}  \rangle =  \prod\limits_{\vec{k}}
(u_{\vec{k}} + v_{\vec{k}}d_{\vec{k}\uparrow}^\dagger d_{-\vec{k}\downarrow}^\dagger)\mid 0 \rangle
\end{eqnarray}
where $d_{\vec k \sigma} = \cos{(\frac{\theta_{\vec k}}{2})} c_{\vec k \sigma} -
\sigma \sin{(\frac{\theta_{\vec k}}{2})} c_{\vec k + \vec Q, \sigma}$, and
 $\vec Q=(\pi, \pi)$ is the magnetic wave vector.
$\prod\limits_i (1 - \alpha n_{i\uparrow}n_{i\downarrow})$ is a
Gutzwiller projection operator, which partially projects out the
doubly occupied electron states on every lattice site and $0 \leq
\alpha \leq 1$ measures the strength of the projection. Obviously,
$\alpha =0$, and $\alpha =1$ correspond to a non-projected and a completely
projected states, respectively. At
$\theta_{\vec k} =0$, we have  $d_{\vec k \sigma}=c_{\vec k
\sigma}$, and $\mid \Psi_{GS} \rangle$ is reduced to a partially projected
BCS state, which we shall loosely call it Gossamer SC state~\cite{laughlin,zhang03}.
In the limit $u_{\vec k}v_{\vec k} =0$, $\mid \Psi_0 \rangle$ is reduced to a
SDW state. The variational parameters are $u_{\vec k}, v_{\vec k}$,
$\theta_{\vec k}$ and $\alpha$. Such a wavefunction should enable us
to study the phase transition between the AFM and SC states. The
metallic or insulating phase can be determined by the continuity of
the chemical potential.

To carry out the variation, we apply the Gutzwiller approximation to
replace the effect of the projection operator by a set of
renormalization factors, which are determined by statistical
countings~\cite{Gutzwiller,vollhardt,zhang88,ogata}. Let $\langle O
\rangle$ be the expectation value of the operator $O$ in the state
$|\Psi_{GS} \rangle$, and $\langle O \rangle_0$ be that in the state
$|\Psi_{0} \rangle$. The Gutzwiller approximation gives
\begin{eqnarray}
\langle c_{i\sigma}^\dagger c_{j\sigma} \rangle = g_t^{ij} \langle
c_{i\sigma}^\dagger c_{j\sigma} \rangle_0  ,\,\,\,
\langle \vec{S}_i\cdot\vec{S}_j \rangle =
g_s \langle \vec{S}_i\cdot\vec{S}_j \rangle_0
\end{eqnarray}
where $g'$s are determined by the ratio of the probability of the corresponding physical processes
in the projected and unprojected states~\cite{zhang88}. We introduce a sublattice
magnetization for sublattices $A$ and $B$,
\begin{eqnarray}
m_0 =\frac{1}{2}\langle n_{A\uparrow} -
n_{A\downarrow} \rangle_0 = - \frac{1}{2}\langle n_{B\uparrow} -n_{B\downarrow} \rangle_0
\end{eqnarray}
$g's$ are then functions of the electron density $n$, $m_0$, and the
double occupation number $d=\langle n_{i\uparrow}n_{i\downarrow}
\rangle$,
\begin{eqnarray}
g_s \, &=& (n-2d)^2/(n-2n_{+}n_{-})^2 \nonumber\\
g^{ij}_t &=& G^iG^j\nonumber \\
G^{ A} &=& g_s^{1/4} [\,s(1-n_-) +\sqrt{n_- d/n_+}\,] \nonumber \\
G^{B} &=& g_s^{1/4} [ \, s(1-n_+) + \sqrt{n_+ d/n_-} \, ]
\end{eqnarray}
In the above equations, $n_{\pm} = \frac{n}{2} \pm m_0 $, and
$s=\sqrt{\frac{1-n+d}{(1-n_+)(1-n_-)}}$. The superindex in $G$
refers to the sublattice of the site. Note that there is a
one-to-one correspondence between $d$ and $\alpha$ given by,
\begin{eqnarray}
1-\alpha= s^2 d/g_sn_{+}n_{-}
\end{eqnarray}
In the absence of the sublattice magnetization, $g_t$ and $g_s$ in
Eqn, (6) are reduced to their values in the uniform state \cite{Gan03},
which are further reduced, in fully projected case ($\alpha=1$ or
$d=0$), to the values in the RVB state~\cite{zhang88}.
Within the Gutzwiller approximation,
the variation of the projected state for $H$ in (1) is reduced
to the variation of the unprojected state $\mid \Psi_0 \rangle$ for
a renormalized Hamiltonian $H_{eff}$,
\begin{eqnarray}
H_{eff} &=& Ud - \sum_{\langle ij \rangle \sigma} g_t^{ij}(c^{\dag}_{i\sigma}c_{j\sigma} + h.c.) \nonumber\\
&& + g_sJ \sum_{(ij)} \vec S_i \cdot \vec S_j -\mu \sum_{i\sigma} n_{i\sigma}
\end{eqnarray}

To proceed further, we introduce a  self-energy $\chi$ and a $d$-wave
pairing amplitude $\Delta$,
\begin{eqnarray}
\chi &=& \sum\limits_\sigma \langle c_{i\sigma}^\dagger c_{i+ \hat x \sigma} \rangle_0
=\sum\limits_\sigma \langle c_{i\sigma}^\dagger c_{i+ \hat y \sigma} \rangle_0 \\
\Delta &=& \sum_{\sigma}\langle \sigma c_{i\sigma}c_{i+ \hat x\,-\sigma} \rangle_0
= -\sum_{\sigma}\langle \sigma c_{i\sigma}c_{i+ \hat y\,-\sigma}\rangle_0
\end{eqnarray}
The singlet SC order parameter $\Delta_{SC} \approx g_{SC} \Delta$, with
$g_{SC} = (g_t^{AA} + g_t^{BB})/2$.
The pairing amplitude and the
SDW state described below defines the variation of $\mid \Psi_0
\rangle$. As in the usual SDW variation, we choose
$\cos{\theta_{\vec k}}=\epsilon_{\vec k}/\xi_{\vec k}$, where
$\epsilon_{\vec k} = -(2tg_t^{AB} +3Jg_s\chi/4)\gamma_{\vec k,+}$ is
the kinetic energy including a self-energy term of $\chi$, and
$\xi_{\vec k} = \sqrt{\epsilon_{\vec k}^2 +
\tilde{\Delta}^2_{AF}(\vec k) }$, with $\tilde{\Delta}_{AF}(\vec k)=
\Delta_{af} + t'(g_t^{AA}-g_t^{BB})\zeta_{\vec k}$. $\Delta_{af}$ is
a variational parameter to determine $m_0$. The second term in
$\tilde{\Delta}_{AF}$ arises from a spin-dependent hopping process
along the [1,1] direction in $H_{eff}$. In the above equations, we
have denoted $\gamma_{\vec k,\pm}=\cos{k_x} \pm \cos{k_y}$, and
$\zeta_{\vec k} = \cos{(k_x + k_y)}$. With the above variational
wavefunction, we calculate the expectation value of $H_{eff}$ and
find the ground state energy,
\begin{eqnarray}
E & = & Ud - 4 g_t t \chi + g_t^{AA}\langle H_{t^\prime}^+
\rangle_0 + g_t^{BB}\langle H_{t^\prime}^- \rangle_0 \nonumber\\
  &   &    - (3g_sJ/4)(\Delta^2 + \chi^2) - 2Jg_sm_0^2
\end{eqnarray}
where $m_0$, $\chi$, $n$, and $\Delta$ are the solutions of their
corresponding self-consistent equations.  The two additional
variational parameters $d$ and $\Delta_{af}$ are to minimalize the
ground state energy. Note that  $ 0 \leq d \leq d_0$, with
$d_0=\langle n_{i\uparrow}n_{i\downarrow} \rangle_0$. In Eqn. (11),
$\langle H_{t^\prime}^{\pm} \rangle$ are given by
\begin{eqnarray}
\langle H_{t^\prime}^{\pm} \rangle_0 & = & -\frac{2t'}{N}\sum_{\vec{k}\in
A} \zeta_{\vec{k}}[v_{\vec{k}}^2 (1 \mp \sin{2\theta_{\vec k}}) +
v_{\vec k+\vec Q}^2 (1 \pm \sin{2\theta_{\vec k}})] \nonumber
\end{eqnarray}
where the summation of $\vec k$ runs over the reduced Brillouin zone, and
\begin{eqnarray}
v_{\vec k}^2 = \frac{1}{2}( 1 - (\xi_{\vec k} - \tilde{\mu})/E^{-}_{\vec k} ),\,\,
v_{\vec k + \vec Q}^2 = \frac{1}{2}( 1 + (\xi_{\vec k} +\tilde{\mu})/E^{+}_{\vec k} ), \nonumber
\end{eqnarray}
with
$E^{\mp}_{\vec k} = \sqrt{(\xi_{\vec k} \mp \tilde{\mu})^2 +\Delta_{\vec k}^2}$, and
$\Delta_{\vec k} = (3/4)Jg_s \Delta \gamma_{\vec k,-}$, \,
$\tilde{\mu} = \mu + t'(g_t^{AA} + g_t^{BB})\zeta_{\vec k}$.

\begin{figure}[t]
\includegraphics[width=6.8cm]{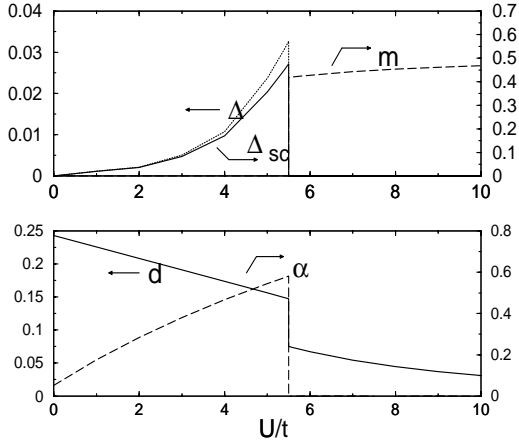}
\vspace{-0.2cm}\caption{\label{Fig1} Pairing amplitude $\Delta$, SC
order parameter $\Delta_{SC}$, and AFM order parameter $m$ (top
panel), and electron double occupancy number $d$ and the projection
parameter $\alpha$ (bottom panel), as functions of $U$ for $J/t=0.5$
and $t^\prime/t=0.8$.}
\end{figure}
We are now ready to discuss our results.  We shall mainly discuss
the half filled case.  At the half filling, there is a critical
$U_c$ to separate a metallic or SC  state at a small $U$ from an AFM
insulator at a large $U$, and the transition is first order with no
co-existence of the two phases. These features are demonstrated in
Fig. 1. There are two regimes in $U$. At $U < U_c (\sim5.5 \, t)$,
$m=0$ while $\Delta$ and $\Delta_{SC}$ increase monotonically as $U$
increases.  $\Delta_{SC}$ is slightly smaller than $\Delta$. This is
a SC state without AFM ordering. At $U > U_c$, $\Delta =
\Delta_{SC}=0$, while $m= {\sqrt g_s} m_0$ changes abruptly from
zero at $U <U_c$ to a saturated value of $0.45$. We have calculated
the chemical potential around the half filling and found it is
discontinuous in the AFM state so that it is an insulator. As we can
see from the bottom panel of Fig. 1, as $U$ increases, $d$ decreases
with a sudden drop at $U_c$ indicating the electron's localization
in the insulating phase, and $\alpha$ increases to its maximum in
the SC phase followed by a discontinuous drop to zero at $U=U_c$.
The latter indicates the absence of the projection in the AFM phase
so that we have $m=m_0$~\cite{projection}. We have also calculated
these quantities with different values of $J/t$ and $t^\prime/t$ and
the results are qualitatively similar except that $\Delta$ becomes
very tiny at smaller $J/t$. Our results are consistent with major
experiments in organic superconductors. As shown in the pressure
experiments~\cite{Lef00}, the phase transition is first order and
the phase boundary between the AFM and SC states merges with the
phase boundary between the insulating and metallic states. In our
theory, the AFM state is always a Mott insulator. Recent NMR
experiments~\cite{Miya95, Miya02} show the proximity of pseudogapped
superconductor and a commensurate AFM ordering with a finite moment
of $0.4 \mu_B$ (or $0.26 \mu_B$) for
$\kappa$-(BEDT-TTF)$_{2}$Cu[N(CN)$_2$]Cl or Br at low temperatures,
which suggests that the magnetic ordering is driven by electron's
strong correlation rather than by the Fermi surface nesting. In the
Gossamer SC state, the quasiparticle energy is governed by
$\Delta$~\cite{zhang88,vannila,Gan03}, which is larger than the SC
order parameter, implying a pseudogap phase~\cite{palee}. The small
difference between $\Delta_{SC}$ and $\Delta$ in our theory is
partly due to the phenomenological model we use, which more favors
AFM state than the Hubbard model does at moderate or large $U$. We
expect the phase boundary in a more accurate theory will be shifted
to the larger $U$ and $\Delta_{SC}/\Delta$ will be smaller.

\begin{figure}[tbp]
\includegraphics[width=7.5cm]{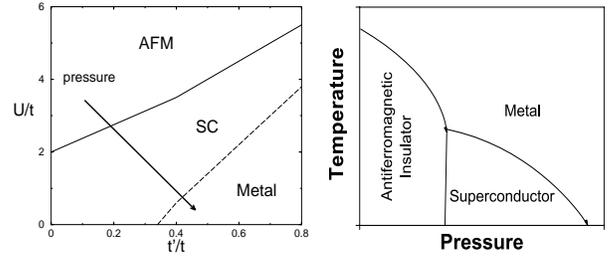}
\caption{\label{Fig2} Left panel: phase diagram of $t'$ v.s. $U$ for
$J/t=0.5$ at the half filling. The arrow indicates the flow of the
parameters under the pressure. Right panel: schematic phase diagram
of organic superconductors.}
\end{figure}
Fig. 2 displays the phase diagram in the parameter space  of $t'$
and $U$ with fixed $J/t=0.5$ at the half filling. There are three
distinct phases. The system is in the AFM phase at large $U$ and
small $t'$, the paramagnetic metallic phase at small $U$ and large
$t'$, and the SC phase at the intermediate parameter region. Here we
have defined a paramagnetic metallic phase if $\Delta \leq 0.01$. At
this very small  $\Delta$, the energy difference between a SC state
and a normal metallic state is practically indistinguishable.  The
phase boundary between the SC and normal states thus obtained is
indicated by a dashed line~\cite{footnote}. For comparison, a
schematic phase diagram abstracted from experimental measurements is
shown at the right panel.  Details of the pressure-temperature phase
diagram of the AFM insulating salt have been
reported~\cite{Lef00,Sasaki02,Muller02}. The effect of pressure in
the schematic phase diagram is to decrease $U/t$ or to increase
$t'/t$.  Our theory is consistent with the general features of this
experimental phase diagram. Note that the $t$-$U$-$J$ model does not
represent the Hubbard model at small $U$. As it is well known, the
ground state of the Hubbard model with $t'=0$ at the half filling is
an antiferromagnet. In our study of Eqn. (1), $J$ is considered as
an independent parameter, so that the $J$-term together with the
kinetic term is in favor of a metallic state at $U=0$ or small
$U/t$.

\begin{figure}[t]
\includegraphics[width=6.8cm]{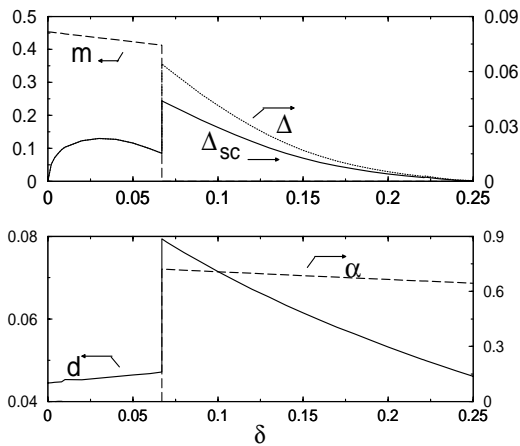}
\vspace{-0.2cm} \caption{\label{Fig3} Doping $\delta$ dependence of
$\Delta$, $\Delta_{SC}$ and $m$ (top panel), and $d$ and $\alpha$
(bottom panel) for $U/t=8$, $J/t=0.5$ and $t^\prime/t=0.8$.}
\end{figure}
Away from the half filling, similar calculations are conducted. The
doping ($\delta=1-n$) dependences of various quantities are plotted
in Fig. 3 for  $U > U_c$.  As $\delta$ increases from zero to a
critical doping $\delta_c \approx 0.07$, the AFM order parameter $m$
decreases slightly, and $\Delta_{SC}$ increases initially, then
saturates.  The ground state is an unprojected SDW and SC state
($\alpha =0$), where the AFM and SC phases co-exist. At $\delta
=\delta_c$,  $m$ drops to zero. The sudden disappearance of the AFM
order strongly enhances the SC pairing. $\Delta_{SC}$ has a jump at
$\delta =\delta_c$ followed by a slow decrease as $\delta$ further
increases. In the region $\delta
> \delta_c$, we have a pure $d$-wave SC state. The essential physics
here is similar to the doped RVB state~\cite{vannila}, except that
here we have a strong first order phase transition on the AFM
ordering at $\delta_c$, a point which requires further study.

In summary, we have presented a strong coupling variational theory
to examine the superconductivity near antiferromagnetic Mott
insulator in layered organic conductors by using a Hubbard model
including a spin-spin coupling term. The theory appears
qualitatively consistent with a number of major experiments, such as
the first order phase transition between AFM Mott insulator and
superconductor under pressure, the merge of the metal-insulator
transition and the AFM-SC transition point, the pseudogapped
phenomenon in the SC state, the large magnetic moment in the AFM
phase, and the transition to the normal metallic phase at high
pressure. The present theory may be further improved by using
variational Monte Carlo calculations on the expectation values and
by developing a more accurate Hamiltonian-trial-wavefunction
approach describing the physics of the layered organic conductors.

This work is partially supported by Chinese Academy of Sciences, and
by the RGC of Hong Kong.


\begin{thebibliography}{99}
\bibitem{organic} D. Jerome, Science {\bf 252}, 1509 (1991).

\bibitem{McKenzie97} R.H. McKenzie, Science {\bf 278}, 820 (1997);
Comments Cond. Mat. Phys. {\bf 18}, 309 (1998).

\bibitem{book1} T. Ishiguro, K. Yamaji, G. Saito: {\itshape Organic
superconductors}, 2nd edn. (Springer, Berlin 1998).

\bibitem{john1} J. Singleton, Rep. Prog. Phys.  {\bf 63}, 1111 (2000).

\bibitem{Lang03}  M. Lang, and J. Mueller, cond-mat/0302157.

\bibitem{Miya95} K. Miyagawa \emph{et al.}, Phys. Rev. Lett. {\bf 75}, 1174 (1995).

\bibitem{Lef00} S. Lefebvre \emph{et al.}, Phys. Rev. Lett. {\bf 85}, 5420 (2000).

\bibitem{Miya02} K. Miyagawa, A. Kawamoto, and K. Kanoda, Phys. Rev. Lett. {\bf 89},
017003 (2002).

\bibitem{Sasaki02} T. Sasaki \emph{et al.}, Phys. Rev. B {\bf 65}, 060505(R)
(2002).

\bibitem{Muller02} J. Muller \emph{et al.}, Phys. Rev. B {\bf 65}, 144521 (2002).

\bibitem{Maya95} H. Mayaffre \emph{et al.}, Phys. Rev. Lett. {\bf 75}, 4122 (1995).

\bibitem{Kanoda} K. Kanoda, Physica C {\bf 282-287}, 299 (1997);
Hyperfine Interactions {\bf 104}, 235 (1997).

\bibitem{arai} T. Arai \emph{et al.}, Phys. Rev. B {\bf 63}, 104518 (2001).

\bibitem{izawa3} K. Izawa \emph{et al.}, Phys. Rev. Lett. {\bf 88}, 027002 (2001).

\bibitem{Kino96} H. Kino and H. Fukuyama, J. Phys. Soc. Jpn. {\bf 65}, 2158 (1996).

\bibitem{Schmalian98} J. Schmalian, Phys. Rev. Lett. {\bf 81}, 4232 (1998).

\bibitem{Kino98} H. Kino and H. Kontani, J. Phys. Soc. Jpn. {\bf 67}, 3691 (1998).

\bibitem{Kondo99} H. Kondo and T. Moriya, J. Phys. Soc. Jpn. {\bf 67}, 3695 (1999);
\emph{ibid.} {\bf 68}, 3170 (1999).

\bibitem{Kino04} H. Kino, H. Kontani, and T. Miyazaki, J. Phys. Soc. Jpn. {\bf 73}, 25 (2004).

\bibitem{Ogata04} Y. Tanaka, Y. Yanase, and M. Ogata, J. Phys. Soc. Jpn. {\bf 73}, 2053 (2004).

\bibitem{Nagaosa} S. Murakami and N. Nagaosa, J. Phys. Soc. Jpn. {\bf 69}, 2395
(2000); S. Onoda and N. Nagaosa, {\em ibid.} {\bf 72}, 2445 (2003).

\bibitem{laughlin} R. Laughlin, cond-mat/0209269.

\bibitem{laughlin2} B.A. Bernevig, R.B. Laughlin, and D.I. Santiago, Phys. Rev. Lett. {\bf
91}, 147003 (2003); B.A. Bernvevig et al. cond-mat/0312573.

\bibitem{zhang03} F.C. Zhang, Phys. Rev. Lett. {\bf 90}, 207002 (2003).

\bibitem{Gan03} J.Y. Gan, F.C. Zhang, and Z.B. Su, cond-mat/0308398.

\bibitem{vannila} P.W. Anderson \emph{et al.}, J. Phys. Cond. Matt. {\bf 24}, R755 (2004).

\bibitem{zhang88} F.C. Zhang \emph{et al.}, Supercond. Sci. and Tech. {\bf 1}, 36 (1988).

\bibitem{randeria} A. Paramekanti, M. Randeria, and N. Trivedi, Phys. Rev. Lett. {\bf 87}, 217002 (2001).

\bibitem{coleman} P. Coleman, Nature {\bf 424}, 625 (2003).

\bibitem{Baskaran} G. Baskaran, Phys. Rev. Lett. {\bf 90}, 197007 (2003).

\bibitem{Lee} T.K. Lee and S. Feng, Phys. Rev. B {\bf 38} 11809 (1988).

\bibitem{Gros} G.J. Chen \emph{et al.}, Phys. Rev. B {\bf 42}, R2662 (1990).

\bibitem{Gutzwiller} M.C. Gutzwiller, Phys. Rev. {\bf 137}, A1726(1965).

\bibitem{vollhardt} D. Vollhardt, Rev. Mod. Phys. {\bf 56}, 99 (1984).

\bibitem{ogata} A. Himeda and M. Ogata, Phys. Rev. B {\bf 60}, R9935 (1999).

\bibitem{projection} At very large $U$, the $J$ term dominates in energy, and the projection
becomes relevant in the SDW state. This value of $U/t$ is about 40 for $J/t=0.5$.

\bibitem{palee}  P.A. Lee and X.G. Wen, Phys. Rev. Lett. {\bf 78}, 4111 (1997).

\bibitem{footnote} The qualitaive feature remains unchanged if we choose different
values for this small number.

\end{thebibliography}
\end{document}